\documentclass{aa}
\usepackage{psfig}
\begin{document}
\thesaurus{ 12
      (02.19.1;
       08.16.7;
       09.02.1;
       09.09.1;
       09.19.2;
       13.25.4) }
\title{The nature of the Vela X-ray "jet"}
\subtitle{The Rayleigh-Taylor instability and the origin of filamentary
structures in the Vela supernova remnant}

\author{Vasilii Gvaramadze\inst{1,2,3}\thanks{{\it Address for
correspondence}: Krasin Str.,
19, ap. 81, Moscow, 123056, Russia; e-mail: vgvaram@mx.iki.rssi.ru}}

\institute{
Abastumani Astrophysical Observatory, Georgian Academy of Sciences, A.Kazbegi ave.
2-a, Tbilisi, 380060, Georgia
\and Sternberg State Astronomical Institute, Universitetskij Prospect 13,
Moscow, 119899, Russia
\and Abdus Salam International Centre for Theoretical Physics, Strada Costiera 11,
P.O. Box 586, 34100 Trieste, Italy
}

\date{Received / accepted }

\maketitle

\begin{abstract}
The nature of the Vela X-ray "jet", recently discovered by
Markwardt \& \"Ogelman (1995), is examined. It is suggested that the
"jet" arises along the interface of domelike deformations of
the Rayleigh-Taylor unstable shell of the Vela supernova
remnant; thereby the "jet"
is interpreted as a part of the general shell of the remnant. The origin of
deformations as well as the general structure of the remnant
are discussed in the
framework of a model based on a cavity explosion of a supernova star. It is
suggested that the shell deformations viewed at various angles
appear as filamentary
structures visible throughout the Vela supernova remnant
at radio, optical, and
X-ray wavelengths. A possible origin of the nebula
of hard X-ray emission detected
by Willmore et al. (1992) around the Vela pulsar is proposed.

\keywords{ Shock waves --
           pulsars: individual: Vela --
           ISM: bubbles --
           ISM: individual objects: Vela supernova remnant --
	   ISM: supernova remnants --
	   X-rays: ISM}

\end{abstract}

%---------------------------------------------------------------------
\section{Introduction}
%---------------------------------------------------------------------
%
Recent observations of the \object{Vela supernova remnant} (SNR)
by the ROSAT X-ray telescope
revealed (\cite{mo}) a filamentary
feature of irregular shape extended for about $45^{\arcmin}$ to the
south-southwest from
the \object{Vela pulsar} position (see also \cite{h}).
The characteristic length and width of this feature were
estimated to be 6.5 and
1.7 pc, respectively, at the pulsar distance of 500 pc. The feature appears at
energies above 0.7 keV (\cite{mo})
and, as it was shown by ASCA observations of Markwardt \&
\"Ogelman (1997; hereafter \cite{mo97} ), 
is visible to energies of at least 7 keV. The end ("head")
of the X-ray feature coincides (\cite{h}; \cite{mo}) with the
center of the brightest radio
component of the Vela SNR, known as \object{Vela\,X}.
Markwardt \& \"Ogelman (1995) interpreted the X-ray filament as a "cocoon"
of hot plasma
enveloping a one-sided jet emanating from the Vela pulsar (see also
Frail et al. 1997 (hereafter \cite{fb}), \cite{mo97}).
In particular, \cite{fb} suggested that the Vela "jet"
represents a channel along
which energy is supplied from the pulsar into the radio source
Vela\,X. Therethrough they supported the proposal of \cite{wp} 
that the Vela\,X is a plerion, i.e. a nebula
powered by the pulsar. \cite{wisi} presented another
view of the nature of the X-ray "jet". According to their scenario, the
pre-supernova system was a binary, and the "jet" is a channel, produced
by a supersonically moving companion star, refilled by the material
of the exploding star. We propose here an alternative explanation for the
origin of the Vela X-ray "jet". We suggest that the "jet"
arises along the interface of domelike deformations of
the shell of the SNR, which is projected by
chance near the line of sight to the Vela pulsar.
This explanation agrees with our (\cite{g8a}) proposal that the
radio source Vela\,X is a part of the general shell of the Vela
SNR (see also \cite{mima}) and implies the similar origin of filamentary
structures visible throughout the remnant in radio, optical and X-ray
ranges. In Sect. 2, we examine the correlation between the "jet" and the
radio source Vela\,X, and discuss the existing models of the Vela X-ray
"jet". In Sect. 3, we briefly review the observational data
relevant to our model of the Vela SNR, which we outline in the same section.
The origin of filamentary structures and the
nature of the X-ray "jet" are considered in Sect. 4. Sect. 5 deals with some
issues related to our model. Sect. 6 summarizes the work.

%---------------------------------------------------------------------
\section{The radio source Vela\,X and the X-ray "jet"}

%---------------------------------------------------------------------

The radio source Vela\,X is the brightest of three main radio components
constituting the Vela SNR (e.g. \cite{ri}, \cite{wb},
\cite{m8}). The maximum of its emission is shifted for about
$45^{\arcmin}$ to the south-southwest from the pulsar.
Radio observations of \cite{m5} and \cite{fb} showed
that a significant part of the radio emission of the Vela\,X appears
as a system of highly polarized ($\simeq 15-20 \%$, in places $>
40 \%$), linear filaments (the characteristic width of which is
about $3^{\arcmin}$),
while the brightest one extends
from the pulsar towards the center of the Vela\,X.
\cite{wp} suggested that the Vela\,X is a plerion.
The main arguments
in support of this point of view
(lately advocated by \cite{dw}, \cite{fb}, and \cite{bt}) are a high degree of
polarization of radio emission at high frequences,
and a flat radio spectrum (these are the distinctive properties of
pulsar-powered nebulae; e.g.
\cite{ws}). Recent observational data allow us to disagree
with the suggestion of \cite{wp}. Our counter-arguments are the
following. The high degree of polarization is not
an exclusive property of the Vela\,X; observations of
\cite{dun} revealed an "arm" of polarization (up to
$50 \%$) running from the Vela\,X site to a latitude of $\simeq
-\,7{\fdg}5$. A comparison of the radio and optical images of
the Vela SNR shows that the polarized "arm" correlates with
optical filaments in the western half of the Vela SNR
(\cite{bg}), whose origin could be attributed to
projection effects in the Rayleigh-Taylor unstable shell of the
remnant (see \cite{g8a} and Sect. 4). This means that
highly polarized structures could be connected not only with
pulsar-powered nebulae, but also with the shell deformations,
whose interaction leads to the compression (i.e. amplification)
of the (regular) magnetic field accumulated in the shell (see
Sect. 3.3).
As regards to the second argument, it was shown by
Gvaramadze (1998a) that discrepancies in the determination of a
spectral index for Vela\,X (\cite{m8}; Weiler \& Panagia 1980; \cite{m0};
\cite{mima}; \cite{ws}; Dwarakanath 1991; Milne 1995)
are connected with smoothing of radio filaments on low-frequency maps,
that leads to low values of the index. This instrumental effect is less
pronounced at high frequencies, and the results obtained in this
case are more reliable. So \cite{m5} showed that the spectral
index of Vela\,X has a value between -0.4 and -0.8, i.e. close
to the spectral indices of two other main radio components of
the Vela SNR (known as Vela\,Y and Vela\,Z), which are equal to
-0.5 (e.g. Dwarakanath 1991).

Another problem of the plerionic interpretation (mentioned by Milne \&
Manchester (1986)) is the non-central location of the Vela\,X.
\cite{fb} believe that the finding of the one-sided pulsar jet solves this
problem. Let us consider their model at length.

FBM\"O found that the brightest radio filament not only stretches from the
pulsar to the center of the Vela\,X, but also outlines the eastern edge of
the X-ray "jet". They mentioned that this is the sole peculiarity of the
brightest filament,
which allows to distinguish it from other ones. At the same time, \cite{fb}
suggested that the brightest filament is a part of a cylindrical
sheet enveloping the X-ray "jet".
They connected the origin of
this sheet with the compression of the ambient magnetic field by a
freely expanding "cocoon" of hot, X-ray-emitting plasma. The
"cocoon" in its turn arises in the course of interaction of a
one-sided jet (originated due to some reason from the pulsar)
with the ambient medium. As a possible observational test for their
model, \cite{fb} proposed that the radio spectral
index of the brightest filament (the "cocoon") should be
different from that of the rest parts of the Vela\,X.
However, this proposal conflicts
with the result of \cite{m5} that the spectral index of
Vela\,X is uniform from filament to filament.

One of the problems of the jet model was pointed out and discussed
by the authors of the model (\cite{fb}, \cite{mo97}). The point is that the
model cannot explain the irregular shape of the "jet" (it is wide close
to the pulsar, then narrows, and then widens again; see Fig. 1 of
\cite{mo}, or Fig. 3 of \cite{fb}). According to the model,
the width of the "jet" should
be maximum close to the pulsar and then monotonically decrease outwards.

As an argument in support of their model, \cite{mo97} mentioned
that only the X-ray "jet"
has a high-temperature ($\simeq 3-4$ keV) component, while
other X-ray filaments in
the Vela SNR have temperatures of $\simeq 0.1-0.2$ keV.
This however conflicts with the observations by \cite{a8}. 
The right panel of Fig. 1 of his paper shows that some
features visible at radio, optical and/or soft X-ray wavelengths
(e.g. the X-ray protrusion labelled by \cite{aet}
as D/D' and the U-type optical filament on the south edge of the
remnant (see e.g. Fig.\ref{1} in the present paper or Fig. 2 in
\cite{g8b})) have obvious hard ($\geq 1.3$ keV) X-ray
counterparts.

Another difficulty of the model is connected with the jet
momentum. \cite{wisi} (see also \cite{kata}) noted that 
the momentum available in the pulsar emission is few
orders of magnitude lower than the required jet momentum.

Also not clear is how the "cocoon" (i.e. a thin cylindrical
sheet enveloping the jet) can emit the X-ray radiation with spectral
characteristics nearly identical to those of a portion of the
SNR at least $30^{\arcmin}$ to the east from the jet (\cite{mo97}).

Wijers \& Sigurdsson (1997) proposed an alternative solution of the "jet"
problem. They suggested that the "jet" is an asymmetrically
expanding bubble filled
with the material from near
the core of the supernova (SN) star.
The expansion of the bubble is powered by the
pulsar spindown energy, while the asymmetry is due to the effect
of the former companion star (a pulsar or a late-type main-sequence star),
which creates a low-density channel in the circumstellar medium.

However, this model also predicts that the "jet" should be wider towards
the Vela pulsar.

The more serious difficulty of the model is that the bubble (the expansion
velocity of which is about $50 \, {\rm km}\,{\rm s}^{-1}$) will never catch a
cavern around the companion star (whose velocity is estimated to be in the
range of $275-400 \, {\rm km}\,{\rm s}^{-1}$). This is easy to see from the
following. The maximum radius of the trailing part of the cavern created by
a companion star (let us choose for example a pulsar) is (e.g. \cite{li})
\begin{displaymath}
R_{\rm cav} ^{\rm max} \,=\,0.03\,{\rm pc}\, \left({\dot{E} \over
{10^{35}\,{\rm erg}\,{\rm s}^{-1}}}\right)^{1/2} \left({n_{\rm ext} \over
{0.16\,{\rm cm}^{-3}}}\right)^{-1/2} \times
\end{displaymath}
\begin{displaymath}
\quad\quad\quad \times \left({T_{\rm ext} \over {0.12\,{\rm
keV}}}\right)^{-1/2} \,\, ,
\end{displaymath}
where $\dot{E}$ is the pulsar spindown luminosity, $n_{\rm ext}$ and $T_{\rm
ext}$ are the number density and the temperature exterior to the bubble,
respectively (the notations and numbers are adopted from
\cite{wisi}). This estimation shows that there is no sufficiently
long, low-density channel
behind the companion star. The situation is even worse if one uses a set
of parameters ($n_{\rm ext} =0.056\,{\rm cm}^{-3} , T_{\rm ext} =4.48\,{\rm
keV})$ from the new model of the Vela "jet" by \cite{mo97}.

The origin of the X-ray "jet" was also discussed in our paper
(\cite{g8a}), where we suggested that the "jet" arised due to the Mach
reflection of two semi-spherical shocks (domelike deformations of the
SNR's shell). This proposal was intended for the explanation of the "jet"
temperature of 1.2 keV given by Markwardt \& \"Ogelman in their first
(1995) paper, but it fails to explain the hard component of the X-ray
emission (\cite{mo97}), and the general appearance of the "jet".

\section{The Vela SNR}

\subsection{General structure}

The Vela SNR is one of the nearest and most extended
SNRs. It has about the same size of $\simeq 7^{\degr} (\simeq 60$ pc)
in radio (Duncan et al. 1996), optical (\cite{pa}) and
X-ray (Aschenbach et al. 1995) ranges. In all
these spectral ranges the Vela SNR appears as a shell source
with a distinct asymmetry along the line perpendicular to the
Galactic plane (see Fig.\ref{f1}). The northeast half of
the main body of the remnant
(faced towards the Galactic plane) has a well determined circular
boundary of radius $\simeq 3{\fdg}5$, whereas the southwest
half is very disordered, with some (X-ray) features extended up to
$\simeq 6{\fdg}5$ from the center (associated with the Vela
pulsar position). There also exist several features
protruding far outside the
northeast rim of the remnant (Aschenbach et al. 1995; \cite{st};
Duncan et al. 1996; Gvaramadze 1998a,b).
The origin of these extended
features and protrusions were considered by Gvaramadze
(1998b) and Bock \& Gvaramadze (1999).
The age of the remnant was estimated to be $\geq
(2-3)\cdot 10^4 $ years (\cite{ly}).

\subsection{The X-ray data}

The first spatially resolved X-ray maps of the Vela SNR (Kahn et al.
1985; hereafter \cite{kg}) confirm the early result
(e.g. \cite{sew}, \cite{g}) that the soft X-ray emission is dominated in the
northwest and southeast quadrants of the remnant, though they also reveal some
limb-brightening on the northeast edge. \cite{kg} found
that the X-ray emission is distributed in a patchy way,
with significant spectral
and intensity variations on angular
scales ranging from several arcminutes to a few degrees.
Another interesting finding of \cite{kg} is that the soft X-ray structures
generally correspond with optical filaments.
Subsequent observations of the Vela SNR reveal (\cite{se}; Aschenbach et al.
1995) that the X-ray extension of this remnant is much larger than that which
was found by \cite{kg}. Particularly, Seward (1990) found a long,
faint arc in the western part of the SNR, which was interpreted as a
rim of the SNR's shell (see also \cite{a3}).
The spectral analysis of the soft X-ray emission showed
(\cite{kg}) that the emission is softer towards the Galactic plane.
The characteristic X-ray temperature was estimated as $\simeq 0.2$ keV,
although a region of hard X-ray emission (with a temperature $\simeq 1.7$
keV) of $1^{\degr}$ radius to the south-southwest of the pulsar was also
discovered. Willmore et al. (1992; hereafter \cite{wesw}) found an elongated
structure of hard X-ray emission (in the 2.5 - 25 keV band)
in the center of the remnant.
This structure stretches nearly symmetrically for
about $1^{\degr}$ on either side of the pulsar in the northeast-southwest
direction. \cite{wesw} suggested 
that this structure is a synchrotron nebula (in
Sect. 5 we discuss an alternative explanation of this structure). Then
Markwardt \& \"Ogelman (1995) discovered the X-ray "jet", which in its turn
correlates with radio (Milne 1995; \cite{fb}) and optical (see
Sect. 4) filaments.
As it was mentioned above, the "jet" extends to the south-southwest from the
pulsar, so it is not aligned with the elongated structure 
found by \cite{wesw}, though
they are partially intersected close to the pulsar. And the last important
result of X-ray studies of the Vela SNR is the discovery of a net of arcs and
loops of intense soft X-rays which covers the whole remnant (\cite{a7}).

It should also be noted that \cite{aet} proposed that the Vela SNR is
immersed in a
large-scale region of hot ($T\simeq 10^6 \, $ K) plasma
(Aschenbach et al. 1995). Such hot surroundings are not unexpected if the
Vela SNR is located inside the \object{Gum Nebula},
which was suggested to be an
old SNR (\cite{re}). We discuss this proposal in Sect. 5 and conclude
that the Vela SNR is rather projected on this region (or vice versa), than
is physically associated with it.

\subsection{Model of the Vela SNR}

The main point of our model of the Vela SNR is the statement
that the general shape
of the Vela SNR might be explained as a result of the
interaction between the SN ejecta/shock and the pre-existing
wind-driven shell created by the SN progenitor star in a
density-stratified interstellar medium
(with a density gradient
perpendicular to the Galactic plane). It is obvious that the
density gradient is not connected with the global stratification
of the gaseous disk of the Galaxy (the scale-height of
the disk few times exceeds the size of the Vela SNR),
but caused by a density inhomogeneity of the local interstellar
medium. Indeed,
the existense of a large-scale region of enhanced density to the
northeast of the Vela SNR (i.e. towards the Galactic plane) follows from the
observations of \cite{da} and \cite{dub}.

Preliminary analysis suggests that the Vela SNR is a result of
type II supernova explosion,
and that a progenitor star was a 15-20 $M_{\odot}$ star with
mechanical luminosity in the
range $L_{\rm w}\, =\, (0.3-3)\cdot 10^{34} \, {\rm
ergs\,s}^{-1}$. This follows from the
fact that though the Vela SNR expands in the low-density ambient
medium, the size of this
remnant is relatively small (as compared with sizes of
wind-driven bubbles and shells
created by more massive and luminous
stars\footnote{See recent observations of such bubbles
by \cite{cb}.}).

The ionizing radiation of the progenitor star creates an H\,II
region, the inner,
homogenized part of which gradually expands due to the
continuous photoevaporation of
density inhomogeneities in stellar environs (\cite{mc4}).
If $L_{\rm w}$ is
much smaller than some characteristic wind luminosity, $L_{\rm
w} ^{\ast} \, \simeq \,
10^{34} \,(S_{46} ^2 /n)^{1/3} \, {\rm egrs\,s}^{-1}$,
where $S_{46}$ is the stellar ionizing
flux in units of $10^{46} \, {\rm photons\,s}^{-1}$ and $n$
is the mean density the ambient
medium would have if were homogenized, the stellar wind flows
through a homogeneous medium
with the number density $\simeq n$ and temperature $T=8000$ K
(for the sake of simplicity we
neglect the density decrease due to the expansion of
the H\,II region). E.g. for B0.5\,V star
with $S_{46} \simeq 7$ (\cite{os}) and
$L_{\rm w} \simeq 10^{34}\, {\rm egrs\,
s}^{-1}$, and for $n=0.1 \, {\rm cm}^{-3}$
(\cite{wasi}, \cite{g}) one has that $L_{\rm w} \ll L_{\rm w} ^{\ast}$,
therefore we can use the self-similar solutions by \cite{ave} and
\cite{wea} to describe the
early evolution of a wind-driven bubble in a homogeneous medium.

Initially the bubble is surrounded by a thin, dense shell of
swept-up interstellar gas of radius
\begin{displaymath}
R_{\rm b} (t) \, = \, 11\,L_{34} ^{1/5} \,
n^{-1/5} \, t_6 ^{3/5} \, {\rm pc} \, \simeq \, 17
\, t_6 ^{3/5} \, {\rm pc} \, \, ,
\end{displaymath}
where $L_{34} = L_{\rm w} /(10^{34} \, {\rm ergs \, s}^{-1}),
t_6 = t/(10^6 \, {\rm years})$,
but eventually the gas pressure in the bubble becomes
comparable to that of the ambient medium, and the bubble stalls, while the
shell disappears. This happens at the moment $
t_{{\rm s},6} \, = \, 0.3\, L_{34} ^{1/2} \, n^{-1/2} \,  \simeq 1 \, ,
$
i.e. $t_{\rm s}$ is more than ten times
smaller than the time spent by the SN progenitor star as
a core hydrogen burning star, which
is $\simeq 1.5\cdot 10^7$ years (see e.g. \cite{van}, \cite{sal}). The radius
of the stalled bubble is
\begin{displaymath}
R_{\rm b} (t_{\rm s} ) \,  = \, R_{\rm s} \, = \,
5.5 \, L_{34} ^{1/2} \, n^{-1/2} \,
{\rm pc} \, \simeq \, 17 \, {\rm pc} \, \, .
\end{displaymath}
Since the star continues to supply the energy in the bubble, the radius of the
bubble continues to grow, $R_{\rm b} (t>t_{\rm s}) =
R_{\rm s} (t/t_{\rm s})^{1/3}$,
until the radiative losses in the bubble interior becomes
comparable to $L_{\rm w}$. It can be shown
that this happens for $t > t_{\rm rad}$,
where $t_{\rm rad} \simeq 10^7$ years.  Then
the bubble recedes to some stable radius $R_{\rm r}$,
at which radiative losses
exactly balance $L_{\rm w}$ (\cite{der}):
\begin{displaymath}
R_{\rm r} \, = \, 2.2 \, L_{34} ^{6/13} n^{-7/13} \,
\simeq \, 8 \, {\rm pc} \, \, .
\end{displaymath}
The time needed for the bubble to shrink to $R_{\rm r}$
is about $(R_{\rm b} (t_{\rm rad} ) -
R_{\rm r} )/c \, \simeq \, 3\cdot 10^6$ years,
where $c=10$ km/s is the isothermal sound speed in the
ambient medium photoionized by the central star (\cite{der}).
The great bulk of radiative losses is connected with a
thin spherical layer close to
$R_{\rm r}$, where nearly the whole mass of the bubble gas
is concentrated. The main body of
the bubble is occupied by a hot, tenuous gas with the number
density $\simeq 10^{-2}
\, n$ (\cite{der}) and temperature $\simeq 100\, T$.
The mass of the bubble gas can be
estimated from the following relation
\begin{displaymath}
M_{\rm b} \, = \, \left({4\pi R_{\rm r} ^3 \over
3}\right)^{8/13} \, \left({L_{\rm w} (nT)^{3/5}
\over \Lambda _{\ast} }\right)^{5/13} \, m_{\rm H} \,
\simeq 0.7\, M_{\odot} \, \, ,
\end{displaymath}
where $\Lambda _{\ast} = 6.2 \cdot 10^{-19} \,\,
{\rm ergs} \, {\rm cm}^{-3} \, {\rm s}^{-1}$ is
a coefficient in the cooling function for the
temperature range $2\cdot 10^5 {\rm K} \leq T
\leq 4\cdot 10^7$ K (see \cite{mc7}), and $m_{\rm H}$
is the mass of a hydrogen atom.

Before a $15 M_{\odot}$ star exploded as a supernova it becomes
for a short time, $t_{\rm RSG}
\simeq 7.5\cdot 10^5$ years, a red supergiant (e.g. \cite{van}).
The H\,II region outside the bubble cools off and
recombines, while the radiative losses in the bubble interior are negligible
on time-scales of $t_{\rm RSG}$. As a result, the bubble
supersonically reexpands in the external
cold medium and creates a new dense shell (see \cite{der};
cf. \cite{shu}).
Let us assume that before the supernova exploded,
the shell expands up to the radius of
$R_{\rm sh} \simeq 30$ pc, i.e. nearly to the current radius
of the Vela SNR. The mean
expansion velocity of the shell is about 30 km/s. The number density
of the shell can be estimated by the formula
\begin{equation}
\label{1}
n_{\rm sh} \, = \, {1\over3}\, {{R_{\rm sh} ^3 -
R_{\rm r} ^3}\over {R_{\rm sh} ^2 h_{\rm sh}}}\, n \,
\, ,
\end{equation}
where $h_{\rm sh}$ is the thickness of the shell. The strength of the magnetic
field accumulated inside the shell can be estimated
from the following relation
\begin{equation}
\label{2}
B_{\rm sh} \, = \, {1\over 2}\,{{R_{\rm sh} ^2 -
R_{\rm r} ^2}\over {R_{\rm sh}
h_{\rm sh}}} \, B \, \, ,
\end{equation}
where $B$ is the strength of the local interstellar magnetic
field (here we neglect, for the sake of
simplicity, the effects connected with the magnetic poles, which
should inevitably
arise in a more realistic approach to the problem
(e.g. \cite{fe})). The thickness of the shell
is determined by the balance of the thermal pressure of the hot bubble
plasma, the magnetic pressure of the shell, and the ram pressure, so we have
\begin{equation}
\label{3}
h_{\rm sh} \, = \, {1 \over 2^{3/2}}\,
{B \over {(4\pi m_{\rm H} n)^{1/2} v_{\rm sh}}}\,
\left(1 - {R_{\rm r} ^2 \over R_{\rm sh} ^2}\right) \, R_{\rm sh} \,\, .
\end{equation}
The strength of the local interstellar magnetic field can be
scaled from the mean Galactic magnetic
field strength $B_{\ast} \simeq 2\cdot 10^{-6}$ G by the following relation
$B \, = \, (n/ n_{\ast})^{2/3} \, B_{\ast}$ ,
where $n_{\ast} = 1\, {\rm cm}^{-3}$ is the mean Galactic number
density, i.e. for $n=0.1 {\rm cm}^{-3}$,
one has $B=0.4\cdot 10^{-6}$ G.
To estimate the parameters of the shell we assume that at the moment of
supernova explosion the shell velocity $v_{\rm sh}$ is equal to 10\,km/s.
From equation (\ref{3}) one has that the shell
thickness is $h_{\rm sh} \simeq 2.7\, {\rm pc}$.
Given $h_{\rm sh}$, we can estimate the number density (see
equation (\ref{1}),
$n_{\rm sh} \simeq 0.4\, {\rm cm}^{-3}$\, ,
and the strength of the magnetic field in the shell
(see equation (\ref{2})),
$B_{\rm sh} \simeq 2\cdot 10^{-6} \, {\rm G}$.
The mass of the shell is
\begin{equation}
\label{4}
M_{\rm sh} \, = \, {4\pi \over 3}\, \left(R_{\rm sh} ^3  -
R_{\rm r} ^3\right) \, m_{\rm H} n \,
\simeq \, 250 M_{\odot} \, \, .
\end{equation}

It is known that the mass of a shell is a decisive
factor, which determines the
evolution of a SN shock (e.g. \cite{fr}). If the mass
of a shell is smaller than $\simeq 50 M_{\rm ej}$, where $M_{\rm ej}$
is the mass of a
SN ejecta, the SN shock
overruns the shell and continues to evolve adiabatically (as
a Sedov-Taylor blast wave). For more massive ones, the SN shock
merges with the shell, and the reaccelerated shell evolves into a
momentum-conserving stage, that is
\begin{equation}
\label{5}
{4\pi \over 3} \, R_{\rm SNR} ^3 (t) m_{\rm H} n v_{\rm SNR} (t) =
{\rm const} \, \, ,
\end{equation}
where $R_{\rm SNR} (t)$ and $v_{\rm SNR} (t)$ are the radius and
the velocity of the
SNR's shell (the former wind-driven shell), respectively. We believe that
just this situation takes place in the Vela SNR. Indeed, a $15 M_{\odot}$
progenitor star ends its evolution as a $\simeq 5\, M_{\odot}$
red supergiant (e.g. \cite{van}), therefore the mass of the supernova ejecta
is $\leq 4 \, M_{\odot}$, i.e. $M_{\rm sh}\, > \,
50 \, M_{\rm ej}$ (see equation (\ref{4})).

It is also known from numerical simulations
(e.g. \cite{tt}) that
the reaccelerated shell acquires a kinetic energy $E_{\rm kin} = \beta E_0$,
where $E_0$ is the initial explosion energy, $\beta = 0.1 - 0.3$.
For the current radius of the Vela SNR $\simeq 32$ pc,
the mass of the shell is
$\simeq 300 \, M_{\sun}$. Assuming that the current
expansion velocity of the Vela
remnant's shell is $\simeq 100 \, {\rm
km}\,{\rm s}^{-1}$ (see \cite{jen})\footnote{Note
that this is the mean velocity of the shell
expansion, whereas some portions of the shell (deformations) expand with much
higher speeds (e.g. \cite{wasi}, \cite{jen}, \cite{dan}).} one has
$E_{\rm kin} \simeq 4\cdot 10^{49} \, {\rm ergs}$,
that corresponds to the initial explosion energy (for $\beta = 0.2$)
$E_0 \simeq 2\cdot 10^{50} \, {\rm ergs}$.

\section{Deformations of the WDS and the origin of filamentary structures}

\subsection{Characteristic scale of deformations}

\begin{figure*}
\psfig{figure=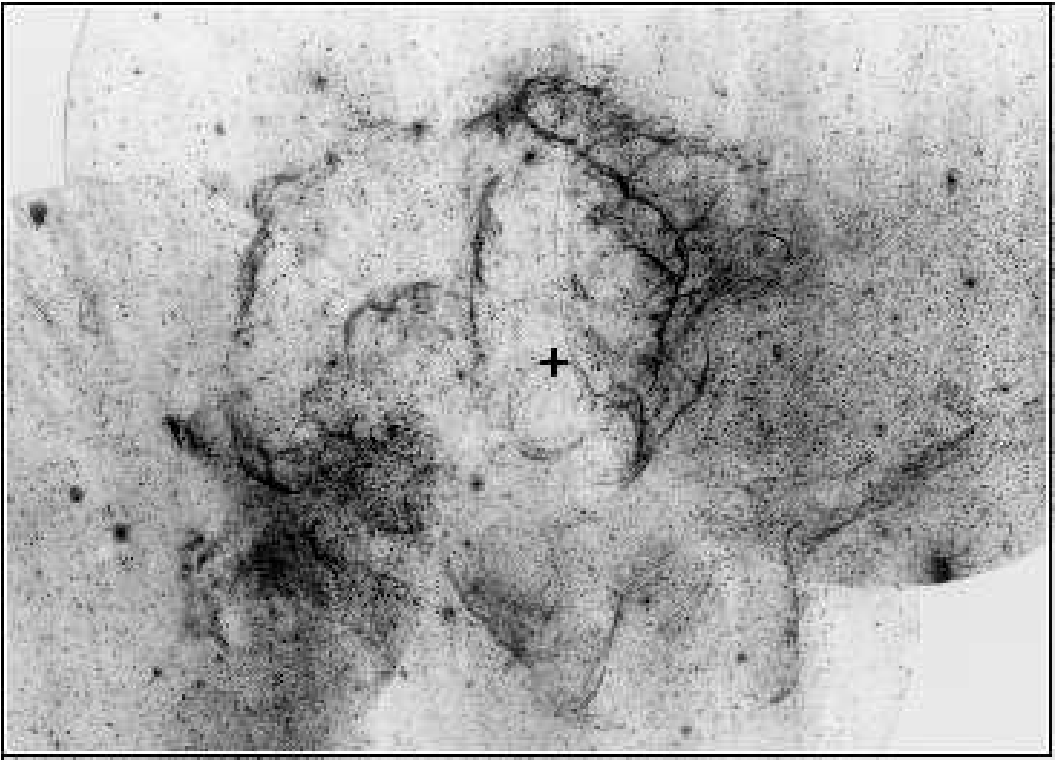,width=14cm}
\caption[]{Mosaic of [O\,III] $ \lambda $ \, 5010 filter photographs of
  the Vela SNR (\cite{pa}). Position of the Vela pulsar is
  indicated by a cross. North is up, east at left.
  The characteristic size of circular filaments
  ("blisters") in the central part of the remnant
  is about $1^{\degr}\, (\simeq 9$ pc).}
\label{f1}
\end{figure*}

After the supernova exploded, the SN ejecta/shock expands almost
freely\footnote{For simplicity, we neglect
the influence of a slow,
dense material lost by the SN progenitor star during the red supergiant
stage on the propagation of the SN shock through the bubble
interior (see however Sect. 5).}
until it catches up with the wind-driven shell.
The impact of the SN ejecta/shock with the shell causes the development
of Rayleigh-Taylor instability (e.g. \cite{sh}).
For the crude order of magnitude estimation of the
characteristic scale $L$ of the
instability (i.e. the characteristic scale of deformations
of the SNR's shell) one
can use the following well-known
formula\footnote{Three-dimensional MHD simulations of the
Rayleigh-Taylor instability developed on the boundary between
two semi-infinite slab regions showed (\cite{ju}) good
correspondence with the predictions of the linear theory.
Analogous numerical simulations but for a finite-thickness
magnetized spherical layer (i.e. a wind-driven shell)
reaccelerated by a blast wave would be highly desirable.} (see
e.g. \cite{bl})
\begin{equation}
\label{6}
L = {2B_{\rm SNR} ^2 \over m_{\rm H} n_{\rm SNR} a} \, \, ,
\end{equation}
where $B_{\rm SNR}$ and $n_{\rm SNR}$ are, respectively, the
magnetic field strength and the number
density in
the reaccelerated wind-driven shell (now the SNR's shell),
$a$ is the acceleration of the shell caused by the
impact of the SN ejecta/shock.
$B_{\rm SNR}$ and $n_{\rm SNR}$ can be estimated by the following formulae
\begin{displaymath}
B_{\rm SNR}  = \left(8\pi m_{\rm H} n\right)^{1/2} \, v_{\rm SNR} (0) \,
\simeq 2.5\cdot 10^{-5} \, \, {\rm G} \, \, ,
\end{displaymath}
\begin{displaymath}
n_{\rm SNR} = {1 \over 3} \, {R_{\rm SNR} \over h_{\rm SNR}} \, n \, \simeq
4.8 \, {\rm cm}^{-3} \, ,
\end{displaymath}
where $v_{\rm SNR} (0) \, \simeq \, 120$ km/s is the
initial expansion velocity of the SNR
(see eq. (\ref{5})),
$R_{\rm SNR}$ is the characteristic radius of the Vela SNR ($\simeq 30$ pc),
$h_{\rm SNR} \, \simeq \,
(B_{\rm sh} /B_{\rm SNR} )h_{\rm sh} \, \simeq 0.2$ pc
is the characteristic thickness of the SNR's shell.
For $a$ one has the estimation
$a = v_{\rm SNR} (0)/\tau$,
where $\tau$ is the crossing time of the SN shock through the
shell (we
assume that the interaction time, i.e. the time of reacceleration, is of
the same order as the crossing time (cf. \cite{cl})),
$\tau = h_{\rm sh} /v' \,$,
and $v'$ is the SN shock velocity in the shell,
$v' \, = \, (6n_{\rm b} / n_{\rm sh} )^{1/2} \, (2E_0 / M_{\rm ej})^{1/2}$.
For the mean number density of the bubble gas, $n_{\rm b} = 3M_{\rm b} /(4\pi
R_{\rm sh} ^3 m_{\rm H}) \, \simeq \,
3\cdot 10^{-4} \, {\rm cm}^{-3}$, and for the mass of the ejecta
$M_{\rm ej} = 3M_{\sun}$ (see Sect. 3.3),
one has $v' \simeq
170 \, {\rm km} \, {\rm s}^{-1}$ and
$a \simeq 2.5\cdot 10^{-5} \, {\rm cm}\,{\rm s}^{-2}$.
Then from (\ref{6}) one has $L \simeq 2 $ pc.
This is a reasonable value, taking
into account the roughness of our analysis.
Note, however, that the scale of deformations depends on the mass (as well as
the magnetic field) distribution over the SNR's shell, and should undergo
considerable changes from place to place.
We see from Fig.\ref{f1} that the instability
initiates small-scale deformations in
the northeast (more massive, see Sect. 3.3) half of the shell,
thereby does not
significantly disturb its initially circular shape. The
situation is more dramatic in the southwest (less massive) half of
the SNR. The dynamical force of the SN ejecta strongly deforms
the shell and even disrupts it in places (see Fig.\ref{f1}).
A hot gas escapes through the gaps in the SNR's shell and forms
radial outflows partially bounded by filamentary structures
(see Gvaramadze (1998a,b),
where an overlay of contours of soft X-ray emission (Aschenbach et
al. 1995) with a mosaic of optical photographs
(\cite{pa}) is presented). The large-scale
domelike deformations of the remnant's
shell, if looked at sideways, appear as U-type or arclike filaments;
numerous examples of such structures exist along the
southeast, south and southwest edges of the Vela SNR. The same
deformations but viewed head-on appear as circular filaments;
two prominent circular filaments are visible in the central part of the
Vela SNR (see Fig.\ref{f1}). We interpret these partially
intersecting filaments as outlines of two "blisters"
on the surface of the SNR, and suggest that these "blisters"
interact with each other due to the lateral
expansion\footnote{The anonymous referee mentioned that due to
the projection effect these blisters may not be physically
adjacent. Though this is not impossible, we believe that the
interaction of blisters really takes place (see Sect. 4.2).}. The
characteristic size of the "blisters" is about $1^{\degr} (\simeq 9$ pc).

\subsection{The interaction of deformations and the origin of filamentary
structures}

The growth of the deformations of the shell is accompanied by the mass
redistribution
over their surfaces. The matter streams down from the
tops of deformations towards
the periphery and accumulates there. The density contrast increases
still more on the nonlinear stage of the Rayleigh-Taylor instability, when
nearby "blisters" come into contact with each other.
The origin of dense sheets of
matter (the so-called spikes) radially
moving with the nearly "free-fall" velocity
(in the reference system of the expanding shell)
is the result of this interaction. The
most dense spikes arise in the regions of collective interaction of several
deformations.

\begin{figure}
\psfig{figure=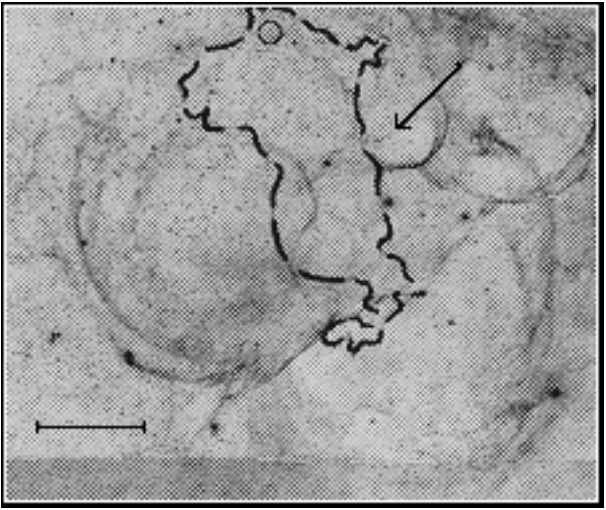,width=8cm}
\caption[]{Two domelike deformations ("blisters") of the Vela SNR's shell
in the central part of the remnant (adopted from \cite{m3}).
The arrow points to the small-scale deformation interacting with
the northwest edge of the eastern "blister". North is up and
east is to the left. The pulsar is shown by a circle. The
contour of low surface brightness of the X-ray "jet" (coded by yellow
colour on Fig. 1 of Markwardt \& \"Ogelman (1995)) is shown by
the dashed line. The horizontal bar is 20 arcmin long.}
\label{f2}
\end{figure}

\begin{figure}
\psfig{figure=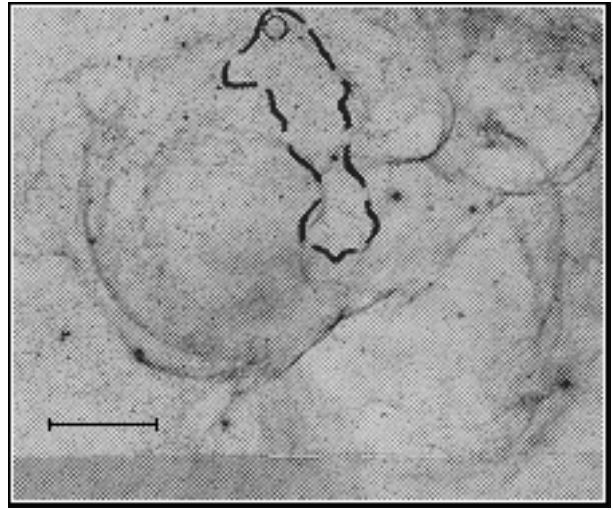,width=8cm}
\caption[]{The same as Fig.\ref{f2}, but for the brightest (red)
portion of the "jet".}
\label{f3}
\end{figure}

The obvious indications of the collective interaction of
deformations exist in the
central part of the
Vela SNR, in particular, to the south and west from the
Vela pulsar (Fig.\ref{f1}; see
also the excellent image of the Vela SNR by \cite{mimu}). Fig.\ref{f2} shows
two circular filaments, which we interpreted in Sect. 4.1 as outlines of two
interacting "blisters" on the SNR's shell. The "blisters"
are surrounded by a few less pronounced deformations of
smaller sizes, one of which
(shown by the arrow) contacts with the northwest edge of the eastern
"blister",
not far from the Vela pulsar position.
Note that the pulsar is situated just on the
(northern) edge of the eastern "blister". The "blisters" meet each other in
the area shifted for about $45^{\arcmin}$ to the south-southwest
from the pulsar. It is of
interest that this area\footnote{We interpret this area as a
Mach shock arising in the course of collision
of two semi-spherical shock waves, or as a splash
of the antispike, which arises on the nonlinear stage of the
Rayleigh-Taylor instability of a thin shell
(see \cite{ot}, \cite{ver}). Our interpretation is
testable. It would be interesting to measure (e.g. through the
absorption lines in spectra of background stars (e.g.
\cite{jen})) the expansion velocities of the shell in the region
of interction of "blisters". One might expect the velocity in
this region to change abruptly (\cite{g8a}), which is
a result of the passage through the surface of the tangential
discontinuity or the boundary of the antispike.},
outlined by distinctly visible optical filaments, coincides both
with the "head" of the X-ray "jet" and with
the center of the radio source Vela\,X.
Moreover, the optical filament on the east edge of the area traces about 60
percent of the X-ray "jet" (see Fig.\ref{f2} and
Fig.\ref{f3}). Another
interesting fact is that most of radio filaments of the radio
source Vela\,X are
lying within the region bounded by two circular filaments
(the outlines of the
"blisters") and showing a fairly good correlation with optical ones
(\cite{g8a}).
These facts suggest that the filamentary structures visible throughout the
Vela SNR in radio, optical and X-ray ranges
have a common nature, and that their
origin is connected with the deformations of the shell.

It is generally accepted that the origin of optical filaments is
connected with the interaction between adiabatic (Sedov-Taylor) SN shocks and
interstellar clouds (\cite{bp}; \cite{mc5}; \cite{de}),
or with projection effects in
radiative shocks, whose fronts are rippled due to the
refraction and reflection by
density inhomogeneities in the ambient medium (\cite{pi}; \cite{so}). The last
mechanism is applicable only to old SNRs,
as it cannot explain the nearly same
optical and X-ray sizes of many remnants. The first one though can explain the
correlation of optical and X-ray emission but fails to explain
the diversity of
sheetlike filaments typical for middle-aged remnants (e.g. the Cygnus Loop).
Instead of this we suggest
that the origin of filaments is connected with projection effects in the
Rayleigh-Taylor unstable shell (the former wind-driven shell).
The shell deformations viewed at
different angles appear (at radio, optical and soft X-ray
ranges) as arclike and looplike
filaments when our line of sight is
tangential to their surfaces. In our model, the optical emission
is expected to come
from the outer layers of the shell, where the transmitted SN
shock slows to become
radiative, while the soft X-ray emission represents the inner
layers of the shell heated by the SN shock up to X-ray temperatures
(cf. \cite{shu}). This explains why the optical
emission outlines so tightly the contours of the soft
X-ray emission, as well as the general correlation of optical
filaments and soft X-ray structures.
However, it should not be one-to-one correspondence of optical and
X-ray filaments. For example, it is quite possible that the less
massive portions of
the SNR's shell were completely overtaken by the transmitted SN
shock, and therefore
are too hot to emit in the optical range. Apparently this is the reason why
we do not
see an optical counterpart to the long, faint arc of soft X-rays
found by Seward (1990) in the west (less massive)
half of the Vela SNR. It is clear that the thickness of
adiabatic parts of the SNR's shell should be of the same
order of magnitude as that of the
former wind-driven shell. The thickness of the arc is
about 2 pc (\cite{a3}), i.e. close to our estimation of
the thickness of the
wind-driven shell, $h_{\rm sh} \simeq 2.7$ pc.

\subsection{Emission measure}

Let us discuss the emission measure of a thermal gas in SNRs.
For normal cosmic abundances it is
\begin{equation}
\label{7}
EM \, = \, \int_{0}^{l} n_{\rm e} ^2 \, dl \,\, ,
\end{equation}
where $n_{\rm e}$ is the electron  number density,
$l$ is a line of sight thickness
of the region occupied by the emitting gas.
It is usually assumed that $n_{\rm e}$
is constant, and, in the
case of a shocked gas, is equal to the immediate postshock density; e.g. for
adiabatic shocks and for a specific heat ratio $\gamma = 5/3 \, ,
n_{\rm e} = 4n$.
The thickness of the postshock emitting region is
$l \, = \, (n/3n_{\rm e} )\, R$ ,
where $R$ is the radius of the shock.
Substituting $n_{\rm e}$ in the last relation,
one has for spherical shock waves
that $l = R/12$. This estimation implies that the X-ray
appearance of Sedov-Taylor remnants should be limb-brightened.

If the gas behind
the shock is in the thermal pressure equilibrium, then
\begin{displaymath}
EM \, \propto \, n_{\rm e} ^2 \, \propto \, T_{\rm s} ^{-2} \, \, ,
\end{displaymath}
where $T_{\rm s}$ is the immediate postshock temperature.
This relation helps to explain the
general softening of the X-ray emission of the Vela SNR
towards the Galactic plane
(i.e. towards the region of enhanced density),
as well as the anticorrelation of
small-scale spectral and intensity variations of the soft X-ray
emission (\cite{kg}). Besides that, the
observable variations of the brightness in regions with the same spectral
characteristics might be explained by changes of $l$ (see (\ref{7})).
Guided by these arguments,
\cite{kg} proposed that the X-ray appearance (in the 0.2-2 keV band)
of the Vela SNR is
connected with the SN shock expansion in a clumpy
medium, and concluded that the progenitor star's ionizing
radiation was too weak to
homogenize the surrounding medium (see \cite{mc4}).
Another conclusion made
by \cite{kg} is that the hard X-ray emission from the region to the south
of the pulsar is nonthermal. This is a consequence of their
assumptions that the emitting gas is in the thermal pressure
equilibrium and that the line of sight extent of the emitting
gas is nearly constant and equal to the diameter of the remnant.
The analysis of the spectral and intensity variations
of the soft X-ray emission of the Vela SNR by \cite{bo} 
was also based on the Sedov-Taylor SN blast wave
model, and they also came to the conclusion that
the SN shock propagates through the clumpy medium.

Our model offers an alternative explanation for the spectral and intensity
variations; such variations arise in a natural way when our line of sight
crosses the deformations of the shell. In particular, the model
explains the origin of numerous arcs and loops observed in soft X-rays
(\cite{a7}). The model could also explain the general
appearance of the remnant in soft X-rays.
As mentioned in Sect. 3.2, the soft X-ray
emission of the Vela SNR does not show limb-brightening
except on the northeast edge.
Our explanation is as follows. The shell deformations result in the increase
of the effective "thickness" of the shell. The "thickening" is
however not uniform around the shell.
The shell deformations are less pronounced on the northeast edge
of the remnant (the more massive, and, correspondingly, the more
stable part of the shell), and we see the limb-brightening
in this direction. The farther from the
Galactic plane, the larger the scale of deformations, and the
larger the "thickness" of the shell. This manifests in two wide
regions of soft X-ray emission in the northwest and southeast directions.

It follows from our model that the main volume of the hot
interior of the Vela SNR does not significantly contribute to
the overall emission
of the remnant (see however Sect. 5). Indeed, substituting $n_{\rm b} \, = \,
3\cdot 10^{-4} \, {\rm cm}^{-3}$ and $l\,\simeq \,60$ pc in (\ref{7}),
one has $EM\, \simeq \,
5.4\cdot 10^{-6} \, {\rm cm}^{-6} \, {\rm pc}$ , i.e. few orders
of magnitude smaller than
the emission measure estimates derived for the Vela SNR's shell
(\cite{kg}) and "jet" (\cite{mo97}).
Therefore we conclude that the hard X-ray emission might be connected with the
innermost layers of the SNR's shell. We suggest that this
emission is thermal and originate in the hot medium evaporated
from the shell, and, in particular, from the dense spikes. We
also suggest that hard X-ray emission should appear as bright
spots in places where spikes penetrate deeply enough
in the hot interior of the remnant to be effectively evaporated
and heated to high temperatures, and where the line of sight
extent of the evaporated gas exceeds that of the gas evaporated
from the adjacent parts of the SNR's shell. Some regular
structures of relatively hard X-ray emission could arise in
regions of collective interaction of deformations. The length
and radial extent of these structures should
be of the same order of magnitude as the scale of deformations, whereas
the width is about the same as the thickness of the shell.
It should be mentioned however that hard X-ray structures should
be less regular than soft X-ray structures since the regions
occupied by the hot evaporated gas are wider than spikes from
which the gas was evaporated. Note that a large number of spots
of hard ($\simeq 2.5 - 10$ keV) X-ray emission was detected
throughout the Vela SNR by \cite{wesw}. We agree with the anonymous
referee that this detection is marginal (only at
3$\sigma$ level or less) and that many of spots may be random
fluctuations. But as it follows from the above discussion, the
possible existence of these spots is not unexpected in
our model. Moreover, our model predicts that spots of hard
X-rays should correlate with soft X-ray structures and optical
filaments. As we already mentioned in Sect. 2, some structures
in the Vela SNR visible at radio, optical and soft X-ray
wavelengths show good correspondence with hard X-ray features.
This correspondence however should not be exact, taking into
account the large radial extent of deformations, and the effect
of projection.

It is known that the magnetic field suppresses the transverse heat
conduction, therefore the magnetic field accumulated in the
shell should decreases
the rate of the mass exchange between the shell
and the hot interior of the SNR. At the same time, one can
expect that the impact of
the SN ejecta/shock with the wind-driven shell
initiates not only large-scale deformations of the
shell, but also small-scale turbulent motions inside the shell
itself, especially in
its innermost layers, where the shell's material experiences the strongest
acceleration. As follows from (\ref{6}), the stronger the
acceleration, the smaller scale of
deformations. \cite{shull} found turbulent motions in the Vela
SNR's shell with
speeds up to $30 \, {\rm km}\,{\rm s}^{-1}$ and the characteristic scale of
deformations $\simeq 10^{-2}$ pc. The turbulent mixing of the
shell's material with the gas in the remnant's interior
leads to the tangling of the magnetic field lines,
and to the subsequent dissipation of the field in the inner
layers of the shell.
This, in its turn, promotes the shell evaporation (see e.g. \cite{cow}).
The reconnection of the magnetic field lines inside the
Rayleigh-Taylor spikes could
be responsible for an additional evaporation of the shell's material.

\subsection{The X-ray "jet" as a dense filament in the shell}

Proceeding from the aforementioned, we suggest that the Vela
X-ray "jet" is a dense filament in the Vela SNR' shell.
There are two main arguments in support of this suggestion.
The first one is that the
X-ray "jet" correlates with the optical and radio filaments,
which should be the
parts of the shell. The second one is that the spectra of the "jet" and its
surroundings (the SNR's shell) are "virtually identical"
(\cite{mo97}). The Vela "jet" is
less than twice brighter than the "background" shell.
One can see from (\ref{7}) that
this enhanced brightness could easily be explained by a small
increase of $l$,
or by an even smaller increase of the local density.
The question arises however why
we do not see many "jets", even not projected on the pulsar?
The possible answer is
that there is no suitable conditions to see them.
The "jets" should appear in areas
which we observe nearly head-on, i.e. in the center of the
remnant; they should be
well-shaped, i.e. should arise in regions of collective
interaction of deformations;
and they should be long enough to be interpreted as "jets", i.e.
the scale of
interacting deformations should be sufficiently large. We see
only one region on the
Vela SNR's surface which could satisfy all of these
requirements, and just this
region is connected with the Vela "jet". However, we do not exclude a
possibility of finding another, less prominent, "jets" in the
region to the west from the pulsar.

Our model does not predict any specific shape of the "jet".
It could be arbitrary,
as its geometry depends only on the freak of chance.
We see that the width of the
"jet" is minimum in the place where the small-scale deformation
of the shell (see Fig.\ref{f2} and Fig.\ref{f3}) interacts with the
northwest edge of the eastern
"blister". We also see that the "jet" has an appendix to the
east from the pulsar. This appendix
has the "right" curvature, taking into account the position of
the pulsar on the
edge of the eastern "blister". From the analysis of the
distribution of radio
filaments over the surface of the radio source Vela\,X we came
to the conclusion
(\cite{g8a}) that the regular magnetic field in the Vela
SNR's shell, if it
exists, should cross the shell along the southeast-northwest
axis (i.e. parallel to the Galactic plane). The magnetic
field introduces an asymmetry in the mass distribution along the
periphery of
domelike deformations (since the matter slides preferentially along the field
lines), and therefore we could expect some
density enhancement on the northwest and southeast outskirts of
the "blisters". The
Vela "jet" lies just on the northwest edge of the eastern "blister", where in
addition to the possible magnetic effect, the collective
interaction of deformations favours its appearance.

M\"O97 showed that the X-ray spectrum of the "head" of the "jet"
can be explained by the
thermal emission of two-component plasma, with a low-temperature component of
temperature $T_{\rm l} \, \simeq \, 0.29$ keV and $EM_{\rm l} \,
\simeq \, 0.28 \, {\rm cm}^{-6} \,$ pc, and a
high-temperature component of temperature $T_{\rm h} \,
\simeq \, 3.74$ keV and $EM_{\rm h} \, \simeq \, 0.18 \, {\rm
cm}^{-6}$ pc. They
also mentioned that these values are indistinguishable, within
the errors, from the
parameters characterizing the emission of the "background"
portion of the SNR located more than $0{\fdg}5$ from the "jet".
Let us adopt these parameters. Then, assuming that the
line of sight
thickness of the "jet" varies from 2 to 9 pc (i.e. it is of the same order of
magnitude as the scale of deformations), one can see that
the number density of the
emitting plasma should be in the range $0.14-0.37 \, {\rm
cm}^{-3}$. Though these estimates are quite reasonable,
they give only a rough idea about the real
parameters of the emitting region, since two of three free parameters for the
fitting of the spectrum (namely, the emission measure and the
temperature) could
suffer significant changes along our line of sight.
An additional (third) thermal
component introduced by \cite{mo97} to describe the predominance
of the "background"
shell emission at energies below 0.7 keV confirms the intricacy
of the problem. Though the fitting of the spectrum is beyond the
scope of our paper, we suggest that the spectrum of the "jet"
should be described by a multi-component thermal model rather
than by a mixed thermal--non-thermal model (M\"O97).

\section{Discussions}

Let us briefly discuss some issues related to our model of the Vela
SNR.

It is worthwhile to note that our model offers a natural explanation for the
"unusual" velocity field inherent to the Vela remnant's shell.
The absorption data by Jenkins et al. (1984)
revealed that the line of sight component of the gas velocity
does not gradually
decrease towards the edges of the remnant (as it should be
according to the standard
Sedov-Taylor model), but rather shows a chaotic behaviour
(see also \cite{jenk}, \cite{dan}). Jenkins et al. (1984)
suggested that some processes should exist which
induce transverse motions in the
shell. The existence of laterally expanding
deformations of the remnant's shell
provides such a process (cf. \cite{me}).

Another issue which has some connection to our model
of the Vela SNR as well as to
the problem of the X-ray "jet" is the origin of the
hard X-ray nebula found by \cite{wesw}. 
As mentioned in Sect. 3.2, \cite{wesw} suggested that
this nebula is powered by the Vela pulsar,
i.e. it is a plerion. Their suggestion was
based on the partial overlapping of the nebula
with the radio source Vela\,X, and on
the spectrum of the nebula, which was fitted
by a power-law model (though it was
stressed that the data do not allow to discern
the thermal and nonthermal forms of
the spectrum). \cite{wesw} estimated the 4-25 keV flux
from the nebula to be $9\cdot
10^{-11} \, {\rm ergs} \, {\rm cm}^{-2} \,
{\rm s}^{-1}$, that at a distance of 500 pc
corresponds to a luminosity of $2.5\cdot 10^{33} \,
{\rm ergs} \, {\rm s}^{-1}$. What
we propose is that the nebula is a dense material lost
by the SN progenitor star during the red
supergiant stage, and reheated to the observed
temperatures after the SN exploded. Let us assume
that this material is homogeneously
dispersed over the whole volume of the nebula, and
shocked to the temperature 10 keV. Then assuming
that the nebula is an oblate spheroid (see Fig.\,1 of \cite{wesw})
with the minor and major semi-axis equal to, respectively,
$0{\fdg}5 \, (\simeq 4.5$ pc) and $1^{\degr} \, (\simeq 9$ pc),
one has that the
number density of the nebula should be $\simeq 0.05 \,
{\rm cm}^{-3} $ to give the
observed flux. This density corresponds to the mass of the emitting gas of
$\simeq 2\, M_{\odot}$, that is a resonable value, taking into
account that the mass
lost by the progenitor star during the red supergiant stage is about
$10 \, M_{\odot}$ (see Sect. 3.3).

In Sect. 3.2 we noted that \cite{aet} proposed that the Vela SNR is
immersed in a large circular area of
X-ray emitting gas. It was suggested that this area
of about $10^{\degr}$ radius
is the hot interior of the Gum Nebula
(which is thought to be an old SNR) and that
the Vela SNR is located inside it. But this proposal
contradicts to the existence of the
optical shell of the Vela SNR and therefore we believe
that the Vela SNR is rather
projected on the ares of hot gas (or vice versa),
than is physically associated with it.

In conclusion we stress that the scenario
presented in this paper for the formation of shell
deformations (i.e. filaments) is valid even if the wind-driven
shell does not exist at the moment of
the SN explosion. In this case, the SN shock wave
hits the wall which bounds a cavity created by the
stellar wind. The density jump at the wall results in the
abrupt deceleration of the SN shock with
the subsequent transition to the radiative stage
of the shock evolution (provided that the column
density of the swept-up
interstellar matter is higher than $\simeq \, 2\cdot 10^{17} \,
v_{100} ^{4.2} \, {\rm cm}^{-2}$,
where $v_{100}$ is the velocity of the transmitted shock
in units of 100 km/s (\cite{mc0})) accompanied by the 
formation of a thin, rippled shell.

\section{Summary}

We have suggested that the Vela X-ray "jet"
arises along the interface of domelike
deformations of the shell of the Vela supernova remnant; thereby the "jet"
has been interpreted as part of the general
remnant's shell. Our suggestion was
based on the comparative analysis of available images of the remnant, and
particularly on the general correlation of filamentary structures visible
throughout the Vela supernova remnant in radio, optical,
and X-ray ranges. We have
proposed that the origin of filaments is connected
with projection effects in the
Rayleigh-Taylor unstable shell; the instability results from the impact of the
supernova ejecta/shock with the pre-existing wind-driven shell created by the
supernova progenitor star. The shell deformations appear
as arclike and looplike
filaments when our line of sight is tangential to their surfaces. The optical
emission is expected to come from the outer layers of the shell, where the
transmitted supernova shock slows to become radiative,
while the soft X-ray emission
represents the inner layers of the shell
heated by the supernova shock up
to X-ray temperatures. The hard X-ray component
is attributed to the evaporately
enhanced medium in the hot interior of the remnant close to the shell. The
nonlinear evolution of the Rayleigh-Taylor
instability results in the formation of
dense spikes, whose deep penetration in the hot
interior of the remnant could lead
to the origin of bright spots of hard X-rays. We have examined the general
appearance of the Vela supernova remnant, and particular attention has been
paid to the soft X-ray structure of the remnant.
A possible origin of the nebula of
hard X-ray emission around the Vela pulsar was also discussed.
It has been suggested
that the nebula is a dense material lost by the
supernova progenitor star during the
red supergiant stage, and reheated after the supernova exploded.

\begin{acknowledgements}
I am grateful to I.Appenzeller and M.Camenzind for their
hospitality during my
stay at the Heidelberg Observatory, where this work was
partially carried out. I am also grateful to B.Aschenbach, 
N.Bochkarev, M.Gilfanov
and J.Tr\"umper for useful discussions, to R.Wijers (the referee) for the
interesting correspondence, and to A.D'Ercole 
for important advices and hospitality
during my visit to the University of Bologna. My thanks also goes to the
anonymous referees, whose
suggestions appreciably changed the paper, to J.Grassberger for assistance in
preparing the figures, and to D.Grilli for carefully reading the
manuscript. This work was partially supported by the
Deutsche Forschungsgemeinshaft (DFG) and the Russian Foundation for Basic
Research (grants: 96-02-00071 and 97-02-16486).

\end{acknowledgements}

\end{document}